\documentclass{aip-cp}

\usepackage[numbers]{natbib}
\usepackage{rotating}
\usepackage{graphicx}
\usepackage{mathrsfs}
\usepackage{amssymb}

\usepackage[normalem]{ulem}
\usepackage{color}
\usepackage{bm}
\usepackage{longtable}
\usepackage{times}

\renewcommand\sout{\bgroup \color{red} \ULdepth=-.5ex \ULset}

\renewcommand{\v}[1]{\textbf{#1}}
\renewcommand{\rm}[1]{\textrm{#1}}

\begin{document}

\title{R-mode instability in compact stars}

\author[aff1]{Cui Zhu}

\author[aff1]{Yu-bin Wang}

\author[aff1,aff2]{Xia Zhou\corref{cor1}}

\affil[aff1]{Xinjiang Astronomical Observatory, Chinese Academy of Sciences, Urumqi, 830011, China.}
\affil[aff2]{Key Laboratory of Radio Astronomy, Chinese Academy of Sciences, Nanjing, 210008, China.}

\corresp[cor1]{Corresponding author: zhouxia@xao.ac.cn}

\maketitle

\begin{abstract}
R-mode oscillations have been identified as viable and promising targets for continuous gravitational wave searches, meanwhile, it would allow us to probe the interior of compact stars directly. As well as emitting gravitational wave, r-modes would strongly affect the thermal and spin evolution of compact stars. In this paper, we reviewed the theory behind the gravitational wave driven r-mode instability in a rapidly rotating compact star. In particular, we will focus on r-mode instability window, r-mode evolution and detectability of r-mode.
\end{abstract}

\section{INTRODUCTION}

The r-mode instability is driven by gravitational radiation via the Chandrasekhar-Friedman-Schutz(CFS) mechanism\cite{Chandrasekhar1970,Friedman1978}, which occurs in all rotating perfect fluid compact stars\cite{ANDERSSON2001}. The mode is generically unstable to the CFS instability, which leads to the rapid growth of the r-mode amplitude. In contrast, the growth of the mode is suppressed by the viscosity of the stellar matter. Given the dependence of gravitational wave
emission on the spin of compact star and the dependence of viscosity on the temperature of a compact star, the critical angular velocity-temperature curve dividing the r-mode stability from instability is introduced in \cite{lindblom_gravitational_1998}. Based on the law of angular momentum conservation, a phenomenological model describing the r-mode evolution was proposed by Owen et al.\cite{owen_gravitational_1998} and improved by Ho \& Lai\cite{Ho2000}. With the development of the observational equipment, such gravitational wave signals might be detected by the advanced Laser Interferometer Gravitational-wave Observatory
(aLIGO) and next generation gravitational-wave observatory\cite{aasi_gravitational_2014,Andersson2018, kokkotas_r-mode_2016}. Former studies were carried out to obtain the properties of the r-mode instability especially for gravitational waves from fast rotating compact stars\cite{Andersson1999,Andersson2000,Andersson2002,sad_quark_nodate,sad_bulk_2007,Andersson2010,Haskell2012, alford_gravitational_2014,Alford2014a, alford_gravitational_2015, moustakidis_effects_2015,haskell_r-modes_2015,mahmoodifar_where_2017,wang_evolution_2017}.

It is realized that the r-modes in a rotating compact star are
generically unstable has led to a considerable effort aimed at understanding the possible astrophysical relevance of this mode. Works concern the interaction between oscillations in the core fluid
and the crust \citep{Andersson2000,Bildsten2000,Lindblom2000,Yoshida2001}, the role of the magnetic field\citep{Spruit1999,Rezzolla2001a,Rezzolla2001b,Cuofano2012}, equation of state(EOS)\citep{moustakidis_effects_2015,Lindblom2000,wen_sensitivity_2012,jaikumar_viscous_2008,idrisy_r_2015,pattnaik_influence_2018}
and the long-term evolution of r-mode\cite{wang_evolution_2017,yu_long-term_2009,alford_impact_2012,cheng_could_2013}. Hydro-dynamical simulations or approximate mode-coupling calculations have been taken out to understand the non-linear saturation of an unstable mode\citep{levin_non-linear_2001,lindblom_numerical_2002,bondarescu_spin_2007,Bondarescu2013,Arras2003}. For a complete set of references, we could see review
articles\citep{ANDERSSON2001,kokkotas_r-mode_2016,haskell_r-modes_2015,Glampedakis2018}.

Restrictive bounds on the
r-mode amplitude were set by electromagnetic observations in recent years\citep{mahmoodifar_where_2017,chugunov_r_2017,jasiulek_r_2017, haskell_are_2017, chirenti_m/r_2018}. After gravitational
waves from the merger of double black holes were detected
by aLIGO\cite{Abbott2016}, we now can test whether or
not such a gravitational wave signal from a compact star can be detectable by gravitational observatories. It is encouraging that despite these strict
limits it turned out that several sources could be above or close to the sensitivity of next-generation detectors and
many dozens of sources should be in reach with the advent of third-generation facilities\citep{aasi_searches_2015}. While emitting gravitational waves, r-mode also would strongly affect the thermal and spin evolution of compact stars\cite{wang_evolution_2017,yu_long-term_2009,cheng_could_2013}. Therefore multi-messenger observations of a spinning compact star can probe the damping of r-mode, which in turn is sensitively dependent on the phases of matter inside the star. To obtain astrophysical constraints and set the theoretical basis of the
r-mode, we will review the theory behind the gravitational wave driven r-mode instability in a rapidly rotating compact star mainly focuses on the r-mode instability window, the r-mode evolution scenarios and the detectability of r-mode. 

\section{ R-mode instability windows}

R-mode is a fluid mode of oscillation for which the restoring force is the Coriolis force, and it only exists in a rotating star. In Newtonian gravity, the mode is characterized by a velocity perturbation $\delta \bf{v}$ in a non-radial direction given in terms of magnetic type vector spherical harmonics $\mathbf{Y}_{\rm{lm}}^{B}$\citep{ANDERSSON2001}:
\begin{equation}
\delta \mathbf{v} = \alpha \left(\frac{r}{R}\right)^l R \Omega \mathbf{Y}_{\rm {lm}}^{B} e^{i \omega t}, 
\end{equation}
where $\alpha$ is the dimensionless amplitude of the mode, $R$ and $\Omega$ are the stellar radius and angular velocity of the unperturbed star, $\omega$ is the frequency of the mode.
This mode is not, however, generically unstable, but only goes unstable above a critical frequency. Although all these modes are generically unstable, the fundamental $l =m = 2$ mode couples most strongly to gravitational waves. At the same time, it dominates the gravitational wave emission\citep{ANDERSSON2001} and is therefore the relevant mode for gravitational wave searches. In the following of this paper, we mainly focus on this $l=m=2$ mode.

In a rotating star, the fluid displacement vector corresponding to the r-mode acquires spheroidal components as well, complicating the mode analysis. However, it is conventional to assume that the modes in rotating stars remain toroidal. It is also accustomed to applying the Cowling approximation in which the back-reaction of the fluid perturbation on the metric, and hence the gravitational potential, is ignored. Previous studies\cite{Provost1981, saio1982} have shown that including such effects modifies the r-mode frequency approximately at the $5\%$ level. Furthermore, if the perturbation is
assumed to be isotropic, $\delta P$ and $\delta \rho$ obeys the same EoS as the unperturbed quantities: P (the pressure) and $\rho$ (the density). With these approximations, the r-mode frequency in the co-rotating frame, to first order in the rotation frequency of the star $\Omega$ is given by\cite{Provost1981}:
\begin{equation}
\omega = \frac{2m\Omega}{l(l+1)} + \mathcal{O}(\Omega^3).
\end{equation}

R-mode evolution is determined by the competition between the viscous damping effect and the
destabilizing effect due to gravitational radiation. The damping timescale associated with a given viscosity mechanism is defined as\cite{lindblom_gravitational_1998}
\begin{equation}\label{tau}
\frac{1}{\tau_{\rm {i}}}=-\frac{1}{2E}\left(\frac{dE}{dt}\right)_{\rm{i}},
\end{equation}
the energy E of the mode is given by \cite{lindblom_gravitational_1998}
\begin{equation}\label{E}
E=\frac{\pi}{2m}(m+1)^3(2m+1)! R^4\Omega^2 \int_0^R dr \rho \left(\frac{r}{R}\right)^{2m+2},
\end{equation}
where we have dropped a proportionality constant that determines the amplitude of the
r-mode, since it cancels in the evaluation of the damping timescale. Explicitly, the gravitational radiation time scale is\cite{lindblom_gravitational_1998}
\begin{equation}
\frac{1}{\tau_{\rm{gw}}}=-\frac{32\pi G\Omega ^{2m+2}}{c^{2m+3}}\frac{(m-1)^{2m}}{[(2m+1)!!]^2}\left(\frac{m+2}{m+1}\right)^{2m+2}\int_0^R dr \rho r^{2m+2},
\end{equation}
while the shear viscosity time scale is\cite{lindblom_gravitational_1998}
\begin{equation}
\frac{1}{\tau_{\rm{sv}}}=\frac{(m-1)(2m+1)}{\int_0^R dr \rho r^{2m+2}} \int_0^R dr\eta r^{2m},
\end{equation}
where $\eta$ is the viscosity coefficient, determined by momentum transport due to different particle scattering.

Bulk viscosity arises because the pressure and density
variations associated with the mode oscillation drive the fluid away from beta
equilibrium\cite{ANDERSSON2001}. It corresponds to an estimate of the extent to which energy is
dissipated from the fluid motion as the weak interaction tries to re-establish
equilibrium. The energy of the mode is lost through bulk viscosity, which is carried away by
neutrinos. The timescales can be calculated by using a simplified expression for the volume
expansion, which is based on approximating the Lagrangian perturbation of the fluid by an
Eulerian one \cite{lindblom_gravitational_1998}
\begin{equation}
\delta\sigma=-i\kappa\Omega\frac{\delta \rho}{\rho},
\end{equation}
this approximation is only good within an order of magnitude for the bulk viscosity damping timescale\cite{Lindblom1999}, the critical angular velocity, which can be constrained by observations, is hardly modified because the bulk viscosity has a steep temperature dependence. The bulk viscosity damping timescale follows:
\begin{equation}
    \frac{1}{\tau_{\rm{bv}}} \equiv -\frac{(dE/dt)_{\rm{bv}}}{E}.
\end{equation}
where the time derivative of the co-rotating frame energy E due to the effect of bulk viscosity is $(dE/dt)_{\v{bv}} = -\int \zeta |\delta\sigma|^2d^3x$. The detail calculation of the bulk viscosity of cold dense matter have been carried out in former works\cite{sad_quark_nodate, sad_bulk_2007, Madsen2000}

The critical rotation frequency $\Omega_{\v{c}}$ of compact stars can be determined by the criterion that at this frequency, the fraction of energy dissipated per unit time exactly cancels the fraction of energy fed into the r-mode by gravitational wave emission\citep{ANDERSSON2001}. That means:
\begin{equation}
\frac{1}{\tau}= \frac{1}{\tau_{\rm{gw}}}+\frac{1}{\tau_{\rm{bv}}}+\frac{1}{\tau_{\rm{sv}}}+\cdots =0,
\end{equation}
where $\tau_{\rm{gw}}$, $\tau_{\rm{sv}}$ and $\tau_{\rm{bv}}$ are gravitational radiation, bulk viscosity and shear viscosity timescales, respectively, and the dots denote other possible dissipated mechanisms. The resulting instability parameter space is commonly depicted as a ``window" in the 
$T-\Omega$ plane and is typically a V-shaped region for a given EOS\citep{ANDERSSON2001,Bildsten2000,chugunov_r_2017,levin_runaway_1999}. 

\begin{figure}
\centerline{%
\includegraphics[width=280pt]{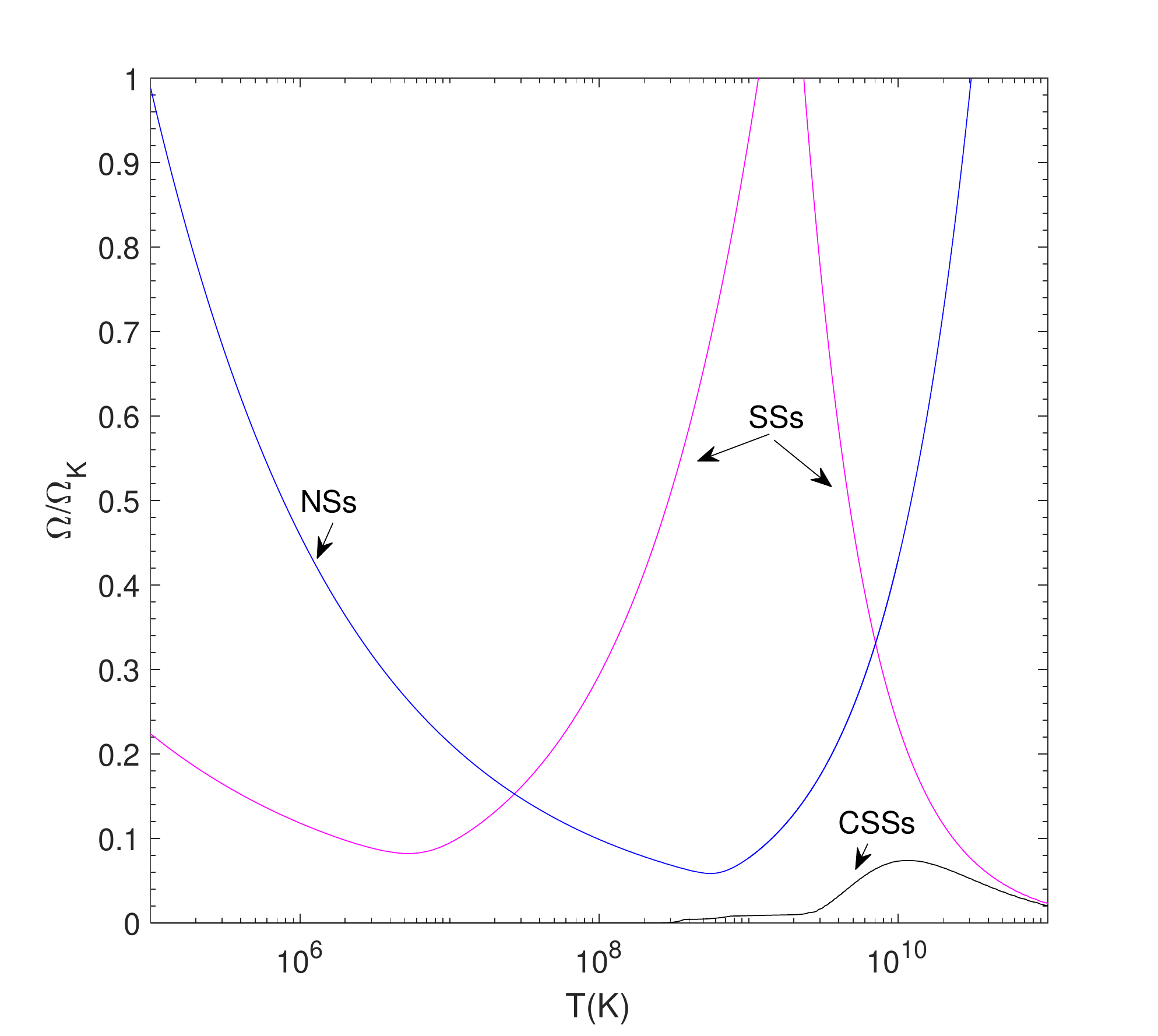}}
\caption{The r-mode instability windows for $1.4M_{\odot}$ neutron stars(NSs, blue line), strange stars(SSs, magenta lines) and CFL phase strange stars(CSSs, black line).}
\label{windows}
\end{figure}

R-modes become only unstable(when $1/\tau<0$) in sources that spin sufficiently fast since viscous dissipation can otherwise damp them away. The boundary of this unstable region corresponds to the points in which the damping and driving timescales are equal. Thus we can get r-mode instability window (or the temperature-dependent critical rotation frequency $\Omega_{\rm{c}}$) by setting $1/\tau=0$. 
In Figure \ref{windows}, we give out the r-mode instability windows for different kinds of $1.4M_{\odot}$ compact star. The instability window of neutron star(NS) is plotted with $n=1$ polytrope. For the strange star(SS) and color-flavor-locked phase strange star(CSS), the corresponding EOSs are described in \cite{Wang2019} . Above the critical frequency, the growth of
the r-modes due to gravitational wave(GW) emission is the most dominant
effect and the star emit GW and lose angular momentum until the effect of viscosities is the dominant one.
 As depicted in Figure \ref{windows}, the region above the temperature-dependent $\Omega_{\rm{c}}$ curve is unstable to r-mode oscillations and the star, if it enters this region, will be spun down rapidly to $\Omega<\Omega_{\rm{c}}$. NS are generally unstable to r-modes except at very high or very low temperatures, where the bulk and shear viscosity respectively are effective at damping
r-mode oscillations. If exotic particles, such as deconfined quarks, are produced in the core
of the NS this generally leads to an increase of the bulk viscosity at low temperatures
which significantly alters the instability window. In Figure \ref{windows}, we also show the instability
window in the case of a SS with $M=1.4M_{\odot}, R=10\rm{km}$. Strange stars made of
ungapped quark matter are more stable than their NS counterparts in a window of
temperatures $10^8 K$ to $10^{10} K$ where the bulk viscosity damping timescale in quark matter
is quite small. CSSs are unstable to r-mode oscillations while $T\sim 10^{10} K-10^{11} K$. 

R-modes are damped by viscous dissipation mechanisms\cite{lindblom_gravitational_1998}, and therefore they connect with the microscopic properties of the matter inside the stars, which depend on the low energy degrees of freedom and EOS, to the macroscopic and observable properties of the star. Most of the uncertainty related to the r-mode instability has to do with its damping mechanisms. Previous studies have tried to constrain the physics behind the r-mode instability of compact stars, especially the EOS of cold dense matter or the additional damping mechanisms(exotic matter\cite{Wang2019}, the viscous Ekman layer\cite{Bildsten2000, Lindblom2000}, Mutual friction\cite{Haskell2009,Gusakov2014} ), by comparing the r-mode instability windows with the observational data of the spin frequency and surface temperature of the compact stars \cite{Haskell2012,chugunov_r_2017,Gusakov2014,Kantor2016,Mahmoodifar2013}. 

\begin{figure}[t]
  \centerline{\includegraphics[width=240pt]{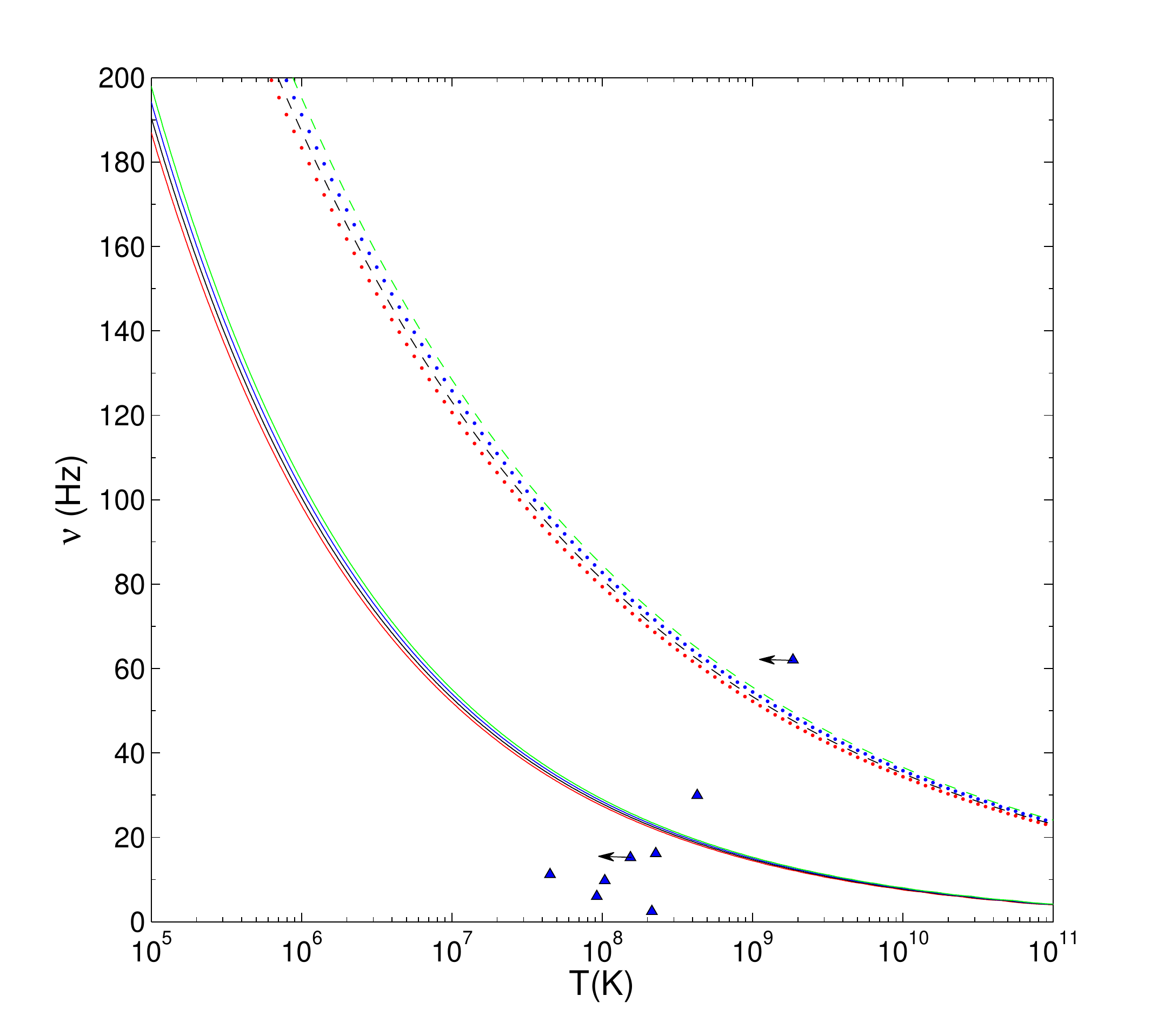}\includegraphics[width=240pt]{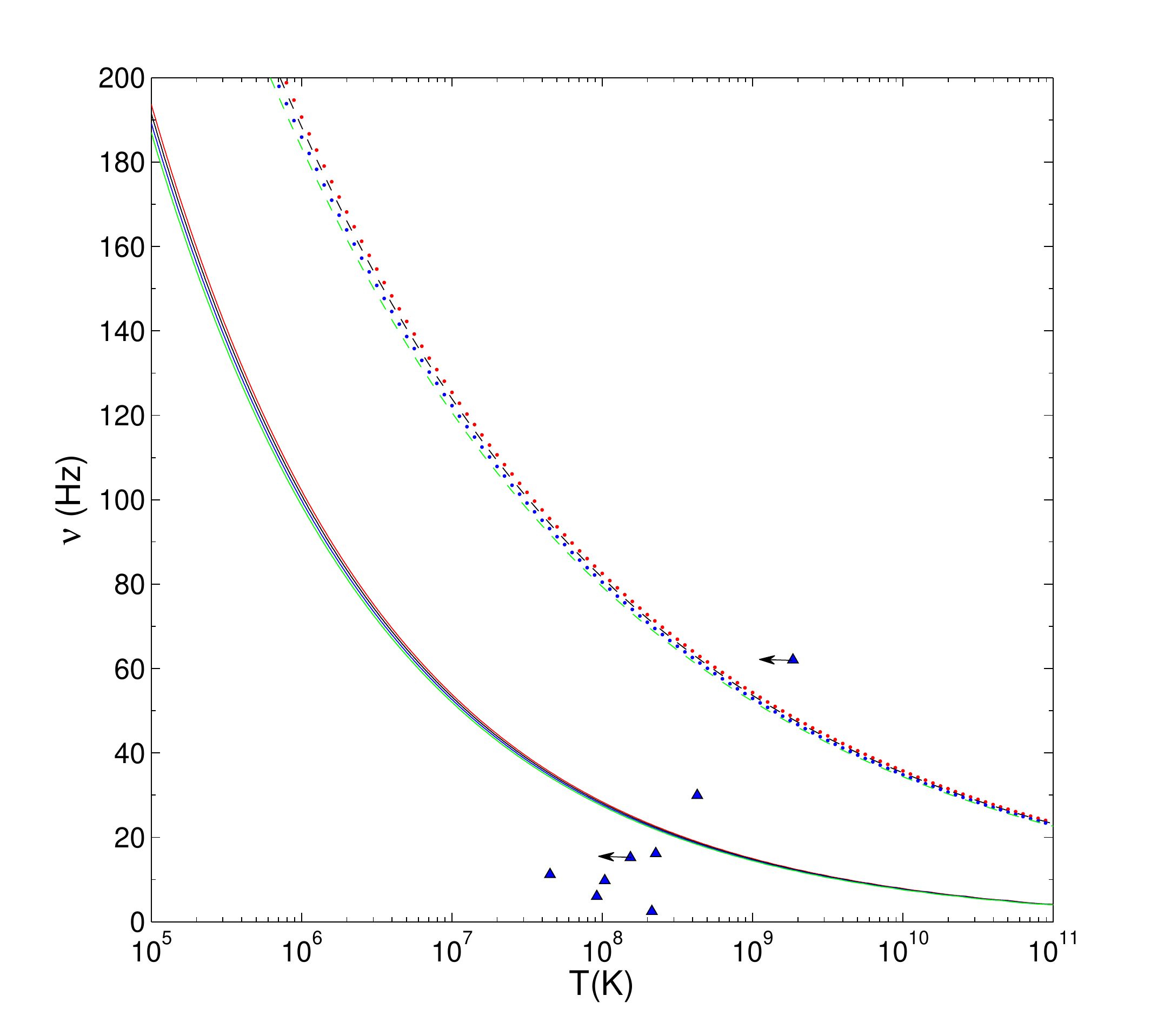}}
  \caption{R-mode instability windows for CSSs with $a_4=1$ and $M=1.4M_{\odot}$. The observed spin frequency and core temperature of young pulsars tabulated in \cite{Wang2019} are also included for comparison(blue triangles). The left panel is for different $B^{1/4}_{\rm{eff}}$ with $\Delta=100\rm{MeV}$, ant the red, black, blue and green lines correspond to $B_{\rm{eff}}^{1/4}=146, 148.5, 151, 153.5 \rm{MeV}$ respectively. The right panel is for different $\Delta$ with $B_{\rm{eff}}^{1/4}=146\rm{MeV}$. The red, black, blue and green lines correspond to $\Delta=70, 80, 90, 100 \rm{MeV}$ respectively. Dotted curves correspond to shear due to surface rubbing, and solid curves correspond to shear due to electron-electron scattering. The figure is taken from \cite{Wang2019}.}
  \label{CFL}
\end{figure}

With the constraints from GW170817\cite{Abbott2017a, Abbott2017b, Zhou2018}, investigations on the r-mode instability window of SSs with unpaired and CFL phase strange quark matter (SQM) are discussed in \cite{Wang2019}, especially the effect of shear viscosity due to the existence of a crust. The r-mode instability windows for CSS with different energy gaps and effective bag constants are given out in Figure \ref{CFL}. The shear due to surface rubbing and electron-electron scattering are considered in this work, it is shown that the suppressed effect of surface rubbing for the instability window is much stronger than that shear by electron-electron scattering. In addition, the viscous timescale for the shear due to the surface rubbing is $\tau_{\rm{sr}}\varpropto 10^9T_{9}$ ($\nu=20 \rm{Hz}$) and the viscous timescale for shear due to electron-electron scattering is $\tau_{\rm{sv}}^{ee} \varpropto 10^{12}T_{9}^{5/3}$. This implies that the dominant role in suppressing the r-modes is supposed to be the shear due to surface rubbing. In Figure \ref{CFL}, most of the young pulsars located out the instability region, except for the fast spinning PSR J0537-6910 and Crab. Whether shear due to surface rubbing or electron-electron scattering is considered, PSR J0537-6910 is always in the instability region. While the star damped by shear due to surface rubbing, Crab is out of the instability windows. This work shows that the effects of EOS of unpaired strange quark matter are only dominated at low temperature, but it do not have a significant effect on the CSSs. For the CSSs, the damping mechanisms due to the existence of the crust are important for us to consider when drawing any conclusions.

As we can see from these works\cite{chugunov_r_2017,Wang2019,Haskell2009,Gusakov2014}, the ``minimal'' model has been described, and that
is often considered in compact star physics, needs to be re-evaluated and additional physics
must be included. And the urgent need for an improved understanding of the various elements in the model of r-mode instability.

\section{R-mode evolution scenarios}

Studying the r-mode evolution of a compact star requires to solve a system of three coupled evolution equations for rotation frequency $\Omega$ of the star and the r-mode amplitude $\alpha$, as well as the temperature $T$. These equations are derived from energy and angular momentum conservation\cite{owen_gravitational_1998,Ho2000}. The total angular momentum $J$ can be quantified as\cite{owen_gravitational_1998}:
\begin{equation}
    J= I\Omega + J_{\rm{c}},
\end{equation}
where $J_{\rm c}$ is the canonical angular momentum of the r-mode,
\begin{equation}
    J_{\v c}=-\frac{3}{2}\alpha^2\tilde{J} M R^2 \omega
\end{equation}
where $\tilde{J}$ is the dimensionless parameters, arising in the canonical energy and angular momentum of the mode. The canonical energy of the
r-mode is given by \cite{owen_gravitational_1998},
\begin{equation}
    E_{\v c}=\frac{1}{2}\alpha^2 \tilde{J}M R^2\Omega^2,
\end{equation}
Then, we obtain the evolution of the r-mode using the time derivative of $J$ and $E_c$\cite{ ANDERSSON2001,owen_gravitational_1998},
\begin{eqnarray}
\frac{dJ}{dt} &=& \frac{3\alpha^2\tilde{J}M R^2 \Omega}{\tau_{\rm{gw}}}+N, \label{jt}\\
\frac{dE_{\rm c}}{dt} &=& -2E_{\rm{c}}\left(\frac{1}{\tau_{\rm{gw}}}+\frac{1}{\tau_{\rm{sv}}}+\frac{1}{\tau_{\rm{bv}}}+\frac{1}{\tau_{\rm{N}}}\right) \label{et},
\end{eqnarray}
where the torque ``N'' stand for other possible mechanisms which affect the total stellar angular momentum and ``$\tau_{\v N}$"stand for corresponding dissipate timescale.

For the ``minimal'' model of r-mode instability, the evolution equations of $\Omega$ and $\alpha$ can be deduced from Equation(\ref{jt}) and (\ref{et}),
\begin{eqnarray}
\frac{d\Omega}{dt}&=&-\frac{\Omega}{\tau_{\rm{m}}}-\frac{2Q\alpha^2\Omega}{\tau_{\rm{v}}},\label{ot}\\
\frac{d\alpha}{dt}&=&\alpha\left(\frac{1}{\tau_{\rm{gw}}}-\frac{1-\alpha^2Q}{\tau_{\rm{v}}}+\frac{1}{2\tau_{\rm{m}}}\right)\label{at},
\end{eqnarray}
where $Q$ is a dimensionless, EOS-dependent parameter, $1/\tau_{\rm{v}}=1/\tau_{\rm{bv}}+1/\tau_{\rm{sv}}$ and $\tau_{\rm{m}}$ is the typical timescale due to magnetic dipole radiation. During the saturation state, Equation(\ref{ot}) and (\ref{at}) would be replaced by
\begin{eqnarray}
\frac{d\Omega}{dt} &=& -\frac{\Omega}{\tau_{\rm{m}}}\frac{1}{1-\kappa Q}-\frac{2\Omega}{\tau_{\rm{gw}}}\frac{\kappa Q}{1-\kappa Q},\\
\frac{d\alpha}{dt} &=& 0,
\end{eqnarray}
where $\kappa=\alpha_{\rm{sat}}^2$ and $\alpha_{\rm{sat}}$ is the saturation amplitude of the r-mode.

Considering the temperature dependence of the viscosities, we would like to show the thermal evolution
equation of a compact star first before calculating the r-mode evolution, which reads as\cite{yu_long-term_2009},
\begin{equation}
C_{\rm v}\frac{d T}{d t}=-L_{\nu}-L_{\gamma}+H,
\end{equation}
where $C_{\rm v}$ is the total specific heat, $L_{\nu}$ is the
total neutrino luminosity and $L_{\gamma}=4{\pi}R^2{\sigma}T_{\rm s}^4$ is the surface photon
luminosity, where $\sigma$ is the Stefan-Boltzmann constant and $T_{\rm s}$ is the
surface temperature and H include
different heating mechanisms, when considering the heating due to r-mode dissipation $H_{\rm{sv}}=(2\alpha \Omega^2 M R^2\tilde{J})/\tau_{\rm{sv}}$.

\begin{figure}[t]
  \centerline{\includegraphics[width=400pt]{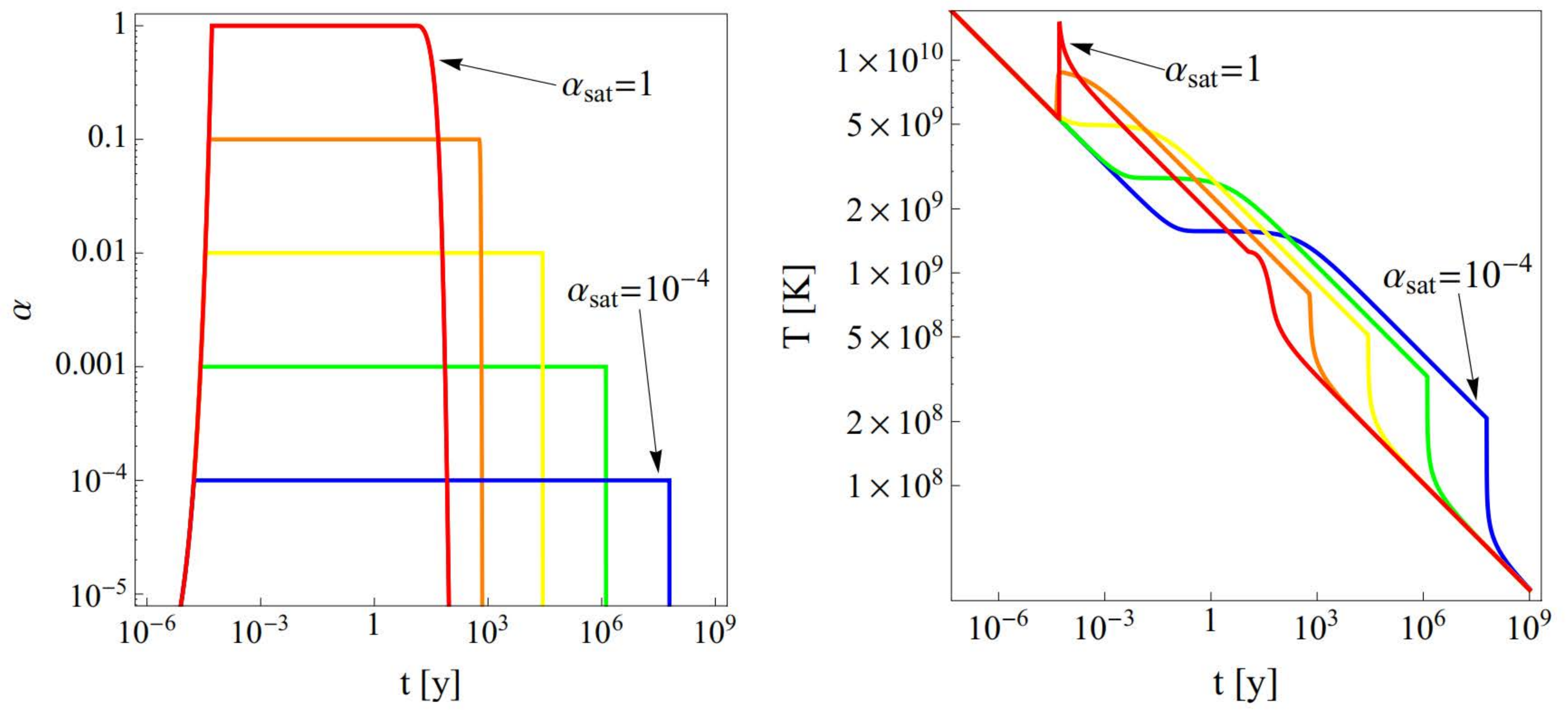}}
  \caption{Evolution of the r-mode amplitude with time(left panel) and evolution of the temperature of the star with time (right panel) for  a young $1.4M_{\odot}$ APR neutron star. The figure is taken from\cite{Alford2012} .}
\label{AT}
\end{figure}

We can see from the above equations that the saturation amplitude $\alpha_{\rm{sat}}$ plays a
critical role in determining the amount of heating and ultimately how far into the
instability window a compact star can move. That means, one of the fundamental importance in judging the astrophysical relevance of the r-mode instability is the determination of saturation amplitude of the mode. On the right panel of Figure (\ref{AT}), the evolution of the temperature of the star versus time is shown. Initially, the r-mode is absent and the star follows a straight line in the logarithmic
$t-T$ diagram corresponding to mere cooling due to neutrino emission. As can be seen, at large
saturation amplitudes $\alpha_{\rm{sat}} \geq 0.1$ the star temporarily cools below the corresponding
steady state curve before the amplitude is large enough so that reheating becomes
relevant. Once the r-mode is saturated the dissipate heating slows the cooling of the
star significantly at all saturation amplitudes. After the star leaves the instability region
and the amplitude goes to zero the star cools further by neutrino emission and approaches
in the logarithmic plot the initial linear behavior, so that the r-mode effectively merely
delays the cooling evolution. The effect of reheating is even larger for smaller values of
$\alpha_{\rm{sat}}$ because in those cases the spin-down takes longer and the star spends more time in
the unstable region, increasing the delay effect.

Theoretical calculations already provide constraints on $\alpha_{\rm{sat}}$ and obviously need to be incorporated in any realistic
r-mode model. In considering the saturation amplitude, its normalization is such that values of order 1 carry energy of the same order of magnitude as the total rotational energy of the compact star. Some authors have introduced damping mechanisms that can cause saturation at r-mode oscillation amplitudes of order $10^{-4}-10^{-2}$ dimensionless units\cite{Bondarescu2009}, while others have introduced mechanisms that cause saturation at amplitudes equal to or larger than $10^{-1}$ \cite{alford_impact_2012}. Another mechanism that may suppress the excitation of the r-mode oscillations is their magneto-hydrodynamic coupling to the stellar magnetic field \cite{Rezzolla2001a, Rezzolla2001b, Cuofano2012, Rezzolla2000}. Moreover, the role of differential rotation in the evolution of the
r-mode was also studied thoroughly \cite{sad_quark_nodate}. While considering millisecond pulsars or low mass X-ray binaries (LMXBs) with measured spin-down rates that reside inside the minimum damping instability window, these data can be used to set upper limits on the r-mode amplitude (obviously, these limits make sense
provided the systems in question are r-mode unstable in the first place, this may not be the case in a more realistic
enhanced damping scenario). The strongest constraints from the available millisecond pulsar or LMXBs data imply a very small amplitude $\alpha\leq 10^{-7}$ \cite{alford_gravitational_2015, mahmoodifar_where_2017, haskell_are_2017}.

All these works introduced in the above paragraphs have obtained a credible conclusion that the saturation amplitude of the r-mode is uncertain. At this point, it is worth pausing to consider the actual gravitational wave detectability of r-mode active compact stars as a function of the mode amplitude and the spin frequency.

\section{R-mode detectability}

The quantitative r-mode evolution scenario discussed in previous paragraphs and works\cite{Andersson2018, Alford2014a, wang_evolution_2017}
provides strong evidence that young pulsars could indeed emit
observable gravitational wave signals. 
Targeted gravitational wave
searches of young neutron stars in supernova remnants can
be adapted to set limits on the amplitude of such oscillations
\cite{Owen2010}. This was first done with a 12-day coherent
search of S5 LIGO data targeting the neutron star in the supernova remnant Cassiopeia A \cite{Wette2008, Abadie2010}. The search has since been extended to nine young supernova remnants, with the most sensitive r-mode
fractional amplitude being less than $4 \times  10^{-5}$
for Vela Jr.  \cite{aasi_gravitational_2014}.
Such a limit is encroaching interesting values of the amplitude when compared to simulations
that calculate the non-linear saturation amplitude of various
r-modes in young compact stars \cite{Bondarescu2009, aasi_gravitational_2014}. Both the steadily growing list of known young pulsars and the strongly increased sensitivity of aLIGO make it a realistic possibility to detect gravitational radiation from
r-mode oscillations in the near future\cite{ligo2019}. 

Not just young pulsars,  but other kinds of compact stars could be promising gravitational wave sources. A detailed analysis of the r-mode detectability can be found in \cite{kokkotas_r-mode_2016}. R-modes become only unstable in sources that spin sufficiently fast since viscous dissipation can otherwise damp them away. It is known that pulsars have a varied evolution: whereas newly formed compact stars are expected to be born with high spin frequencies due to the dramatic spin up during core collapse, the observed young pulsars all have rather low spin frequencies but still large spin-down rates. Generally, during their evolution, they spin down
further and their radio emission stops once their spin-down power becomes too small to power the pulsar jets. However, when compact stars happen to be in a binary, they can be recycled by accretion spin-up, and being heated by accretion and subsequent nuclear reactions they emit
thermal X-ray radiation. Sources in such LMXBs reach frequencies of many hundreds of
Hertz. Once the accretion stops they are not strongly heated any more and turn into stable millisecond pulsars with highly precise observed timing data and which hardly spin down. Based on this general evolution, Kokkotas \& Schwenzer \cite{kokkotas_r-mode_2016} suggested that four classes of sources in which r-modes could
be unstable, namely,(1) newly formed (proto-)compact stars in the immediate
aftermath of a supernova explosion or merger event;
(2) young sources, under a thousand years old, that are
still in their initial rapid spin down;
(3) sources in LMXBs that accrete and on average spin up;
(4) millisecond pulsars that can have ages of billions of
years and only very slowly spin down.

\begin{figure}
\centerline{\includegraphics[width=350pt]{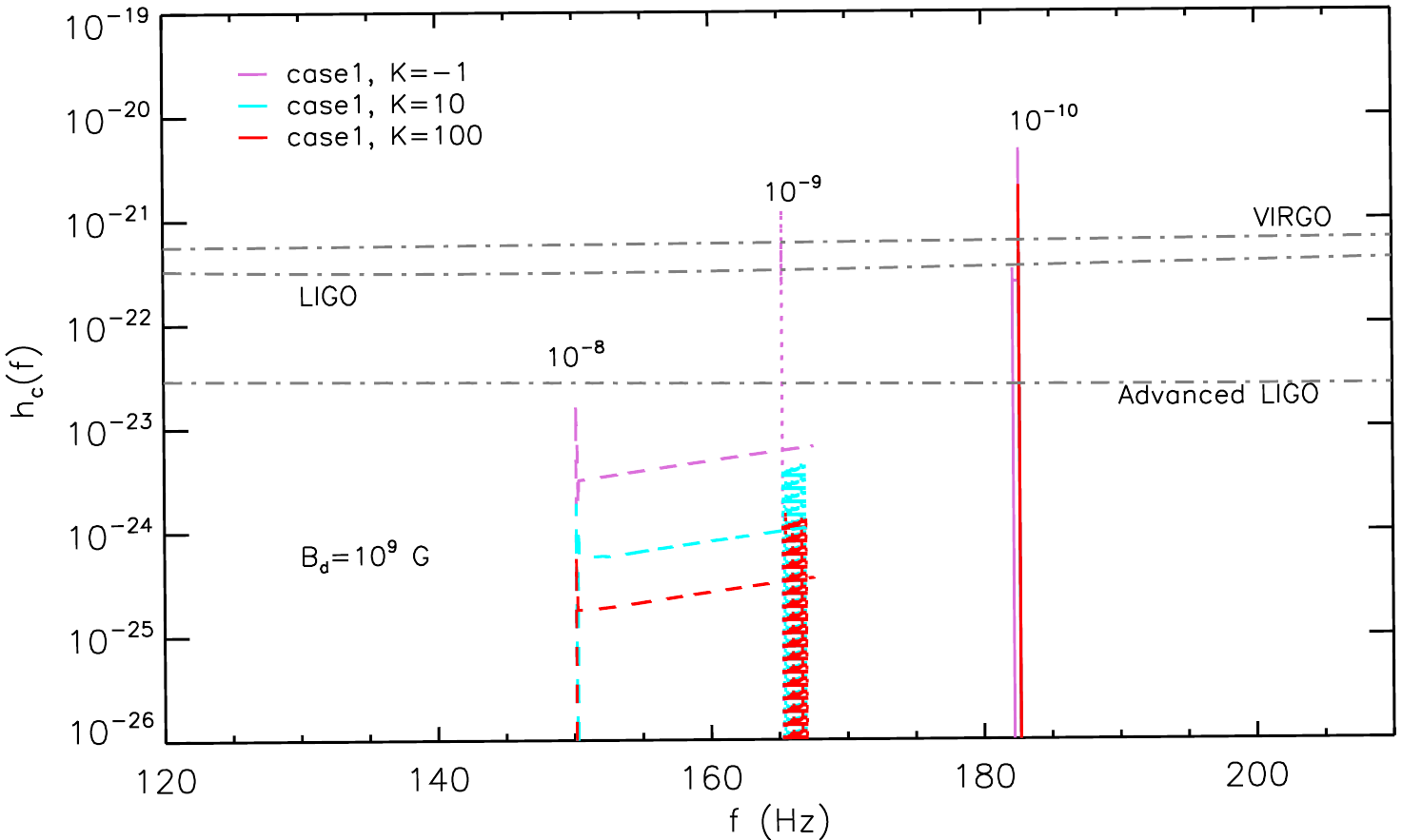}}
\caption{Characteristic amplitude of the gravitational waves $h_{\v c}$ compared with the rms strain noise (dash-dot lines) in the detectors of LIGO, VIRGO and advanced LIGO. Dashed lines correspond to the situation where the mass accretion rate is $10^{-8}~M_{\odot}~yr^{-1}$, dotted lines refer to $10^{-9}M_{\odot}~yr^{-1}$ and solid lines refer to $10^{-10}M_{\odot}yr^{-1}$. Here, $K=-1$ (pink line), $K=10$ ( indigo line) and $K=100$ (red line). The figure is taken from \cite{cao2015}.}
\label{fig:f_hc}
\end{figure}

Since compact stars are perfect point sources, their gravitational wave emission can be described within a multipole
expansion. Here we give the expressions for the r-mode
current quadrupole moment and the relevant derived expressions in a slightly generalized form that takes
into account a general r-mode spectrum parametrized by
the function $\kappa(\Omega)$ in $\omega=(\kappa(\Omega)-2)\Omega\equiv -\frac{4}{3}\kappa(\Omega)\Omega$.
The detectability of gravitational waves in a terrestrial
detector is standard described in terms of the intrinsic gravitational wave strain amplitude $h$. In the case
of $l=m=2$ r-modes it is obtained from the quadrupole expression\cite{owen_gravitational_1998}
\begin{equation}
|h(t)|=\frac{256}{135}\sqrt{\frac{\pi}{30}}\frac{GM}{c^5}\frac{\alpha \Omega^3 R^3}{D}\tilde{J}
\label{hct}
\end{equation}
where $c$ is the speed of light. A characteristic amplitude $h_{\v c} = h\sqrt{f^2 |\frac{dt}{df}|}$. 
In the work \cite{cao2015}, we have studied the role of magnetic damping in the differential rotation of nonlinear r-modes of accreting neutron stars by taking into account the magnetic damping. The results show that the magnetic damping suppresses the nonlinear evolution of r-modes since the saturation amplitude is reduced to a great extent. In particular, due to the generation of a toroidal magnetic field, the spin-down of neutron stars can be stopped, meanwhile, the viscous heating effects are weakened. We may obtain a stronger generated toroidal magnetic field in the framework of the second-order r-mode theory. We also conclude that the detectability of GW from the r-mode instability of accreting neutron stars is drastically reduced as the initial value of differential rotation increases and the accreting rate increases, which are shown in Figure \ref{fig:f_hc}. With the influence of generated toroidal magnetic field on the differential rotation of nonlinear r-modes, GW from accreting neutron stars could be detected by aLIGO detectors if the amount of differential rotation is small when the r-mode instability becomes active and the mass accretion rate is not very high.

For realistic estimates on the chances to use the continuous r-mode emission from compact stars for gravitational wave astronomy, it is crucial to determine the
strength of these signals for the different classes
of sources. The detectability of gravitational waves from
r-modes depends on the gravitational wave amplitude which according to Equation(\ref{hct}) is directly proportional
to the r-mode amplitude $\alpha$ and also depends on the frequency and the distance of a given source, which is also connected with different damping mechanisms.

\section{CONCLUSION}

In this paper, we have reviewed the theoretical aspects of the r-mode instability in a rotating compact star. We also described what the instability window, r-mode evolution and r-mode detectable is predicted to be in a compact star. The predictions of the model can be compared to observations of electromagnetic observations and gravitational detection upper limit. We conclude that r-mode astronomy presents a unique way to study properties of compact stars ranging from bulk parameters to the composition of their opaque interior, and would provide us with precision information that would in many cases be virtually impossible to obtain otherwise. Future next-generation gravitational wave detectors, like the currently planned Einstein telescope\cite{Punturo2010}, should then allow us to see many more potential sources at larger distances. And the combination of the data from next-generation gravitational wave detectors with multi-band electromagnetic observations(like FAST\cite{Li2018a}, QTT\cite{Wang2017}, SKA\cite{Watts2015}, HXMT\cite{Li2018}, NICER\cite{Keek2018}, eXTP\cite{Zhang2019} et al.) could give us a deep inside into the internal of compact stars.

\section{ACKNOWLEDGMENTS}
The work was supported by the West Light Foundation of Chinese
Academy of Sciences (No. 2018-XBQNXZ-B-025) and the
National Natural Science Foundation of China (Nos.11873040, 11373006 )



\begin{thebibliography}{100}  

\bibitem{Chandrasekhar1970} S. Chandrasekhar, Phys. Rev. Lett. \textbf{24}, 611 (1970).

\bibitem{Friedman1978} J. L. Friedman and B. F. Schutz, Astrophys. J. \textbf{221}, 937 (1978).

\bibitem{ANDERSSON2001} N. Andersson and K. D. Kokkotas, Int. J. Mod. Phys. D \textbf{10}, 381 (2001).

\bibitem{lindblom_gravitational_1998} L. Lindblom, B. J. Owen and S. M. Morsink, Phys. Rev. Lett. \textbf{80}, 4843 (1998).

\bibitem{owen_gravitational_1998} B. J. Owen, L. Lindblom, C. Cutler, B. F. Schutz, and N. Andersson, Phys. Rev. D \textbf{58}, 084020 (1998).

\bibitem{Ho2000} W. C. G. Ho and D. Lai, Astrophys. J. \textbf{543}, 386 (2000).

\bibitem{aasi_gravitational_2014} J. Aasi, et al. Astrophys. J. \textbf{785}, 119 (2014).

\bibitem{Andersson2018} N. Andersson, et al. Astrophys. J. \textbf{864}, 137 (2018).

\bibitem{kokkotas_r-mode_2016} K. D. Kokkotas and K. Schwenzer, Eur. Phys. J. A \textbf{52}, 15 (2016).

\bibitem{Andersson1999} N. Andersson, K. D. Kokkotas, and N. Stergioulas, Astrophys. J. \textbf{516}, 307 (1999).

\bibitem{Andersson2000} N. Andersson, D. I. Jones, K. D. Kokkotas, and N. Stergioulas, Astrophys. J. \textbf{534}, L75 (2000).

\bibitem{Andersson2002} N. Andersson, D. I. Jones, and K. D. Kokkotas, Mon. Not. R. Astron. Soc. \textbf{337}, 1224 (2002).

\bibitem{sad_quark_nodate} P. M. S$\acute{a}$, B. Tom$\acute{e}$, Phys. Rev. D \textbf{71}, 044007 (2005).

\bibitem{sad_bulk_2007} B. A. S$\acute{a}$, I. A.  Shovkovy and D. H. Rischke, Phys. Rev. D \textbf{75}, 125004 (2007).

\bibitem{Andersson2010} N. Andersson, B. Haskell and G. L. Comer, Phys. Rev. D \textbf{82}, 023007 (2010).

\bibitem{Haskell2012} B. Haskell, N. Degenaar, and W. C. G. Ho, Mon. Not. Roy. Astron. Soc. \textbf{424}, 93 (2012).

\bibitem{alford_gravitational_2014} M. G. Alford and K. Schwenzer, Phys. Rev. Lett. \textbf{113}, 251102 (2014).

\bibitem{Alford2014a} M. G. Alford and K. Schwenzer, Astrophys. J. \textbf{781}, 22 (2014).

\bibitem{alford_gravitational_2015} M. G. Alford and K. Schwenzer, Mon. Not. R. Astron. Soc. \textbf{446}, 3631 (2015).

\bibitem{moustakidis_effects_2015} C. C. Moustakidis, Phys. Rev. C \textbf{91}, 035804 (2015).

\bibitem{haskell_r-modes_2015} B. Haskell, Int. Journal of Mod. Phys. E \textbf{24}, 1541007 (2015).

\bibitem{mahmoodifar_where_2017} S. Mahmoodifar and T. Strohmayer, Astrophys. J. \textbf{840}, 1538 (2017).

\bibitem{wang_evolution_2017} J. S. Wang and Z. G. Dai, Astron. Astrophys. \textbf{603}, 1432 (2017).

\bibitem{Bildsten2000} L. Bildsten and G. Ushomirsky, Astrophys. J. \textbf{529}, L33 (2000).

\bibitem{Lindblom2000} L. Lindblom, B. J. Owen, and G. Ushomirsky, Phys. Rev. D \textbf{62}, 084030 (2000).

\bibitem{Yoshida2001} S. Yoshida and U. Lee, Astrophys. J. \textbf{546}, 1121 (2001).

\bibitem{Spruit1999} H. C. Spruit, Astron. Astrophys. \textbf{349}, 189 (1999).

\bibitem{Rezzolla2001a} L. Rezzolla, F. K. Lamb, D. Markovic, and S. L. Shapiro, Phys. Rev. D \textbf{64}, 104013 (2001).

\bibitem{Rezzolla2001b} L. Rezzolla, F. K. Lamb, D. Markovic, and S. L. Shapiro, Phys. Rev. D \textbf{64}, 104014 (2001).

\bibitem{Cuofano2012} C. Cuofano, S. Dall’Osso, A. Drago, and L. Stella, Phys. Rev. D \textbf{86}, 044004 (2012).

\bibitem{wen_sensitivity_2012} D. H. Wen, W. G. Newton and B. A. Li, Phys. Rev. C \textbf{85}, 025801 (2012).

\bibitem{jaikumar_viscous_2008} P. Jaikumar, G. Rupak and A. W. Steiner, Phys. Rev. D \textbf{78}, 123007 (2008).

\bibitem{idrisy_r_2015} A. Idrisy, B. J. Owen, and D. I. Jones, Phys. Rev. D \textbf{91}, 024001 (2015).

\bibitem{pattnaik_influence_2018} S. P. Pattnaik et al., J. Phys. G \textbf{45}, 055202 (2018).

\bibitem{yu_long-term_2009} Y. W. Yu, X. F. Cao and X. P. Zheng, Research in Astron. Astrophys. \textbf{9}, 1024 (2009).

\bibitem{alford_impact_2012} M. G. Alford, S. Mahmoodifar and K. Schwenzer, Phys. Rev. D \textbf{85}, 044051 (2012).

\bibitem{cheng_could_2013} Q. Cheng, Y. W. Yu and X. P. Zheng, Phys. Rev. D \textbf{87}, 063009 (2013).

\bibitem{levin_non-linear_2001} Y. Levin and G. Ushomirsky, Mon. Not. R. Astron. Soc. \textbf{324}, 917 (2001).

\bibitem{lindblom_numerical_2002} L. Lindblom and B. J. Owen, Phys. Rev. D \textbf{65}, 063006 (2002).

\bibitem{bondarescu_spin_2007} R. Bondarescu, S. A. Teukolsky, and I. Wasserman, Phys. Rev. D \textbf{76}, 064019 (2007).

\bibitem{Bondarescu2013} R. Bondarescu and I. Wasserman, Astrophys. J. \textbf{778}, 9 (2013).

\bibitem{Arras2003} P. Arras, E. E. Flanagan, S. M. Morsink, A. K. Schenk, T. S. A., and I. Wasserman, Astrophys. J. \textbf{591}, 1129 (2003).

\bibitem{Glampedakis2018} K. Glampedakis and L. Gualtieri, Astrophysics and Space Science Library \textbf{457}, 673 (2018).

\bibitem{chugunov_r_2017}  A. I. Chugunov, M. E. Gusakov and E. M. Kantor, Mon. Not. R. Astron. Soc. \textbf{468}, 291 (2017).

\bibitem{jasiulek_r_2017} M. Jasiulek and C. Chirenti, Phys. Rev. D \textbf{95}, 064060 (2017).

\bibitem{haskell_are_2017} B. Haskell and A. Patruno, Phys. Rev. Lett. \textbf{119}, 161103 (2017).

\bibitem{chirenti_m/r_2018} C. Chirenti and M. Jasiulek, Mon. Not. R. Astron. Soc. \textbf{476}, 354 (2018).

\bibitem{Abbott2016} B. P. Abbott, et al. Phys. Rev. Lett., \textbf{116}, 061102 (2016).

\bibitem{aasi_searches_2015} J. Aasi, et al., Astrophys. J. \textbf{813}, 16 (2015).

\bibitem{Provost1981} J. Provost, G. Berthomeiu and A. Rocca, Living Reviews in Relativity \textbf{94}, 126 (1981).

\bibitem{saio1982} H. Saio, Astrophys. J. \textbf{256}, 717 (1982).

\bibitem{Lindblom1999} L. Lindblom, G. Mendell and B. J. Owen, Phys. Rev. D \textbf{60}, 064006 (1999).

\bibitem{Madsen2000} J. Madsen, Phys. Rev. Lett. \textbf{85}, 10 (2000).

\bibitem{levin_runaway_1999} Y. Levin, Astrophys. J. \textbf{517}, 328 (1999).

\bibitem{Wang2019} Y. B. Wang, et al., Research in Astron. Astrophys. \textbf{19}, 30 (2019).

\bibitem{Haskell2009} B. Haskell, N. Andersson and A. Passamonti, Mon. Not. R. Astron. Soc. \textbf{397}, 1464 (2009).

\bibitem{Gusakov2014} M. E. Gusakov, A. I. Chugunov, and E. M. Kantor, Phys. Rev. Lett. \textbf{112}, 151101 (2014).

\bibitem{Kantor2016} E. M. Kantor and M. E. Gusakov, Mon. Not. Roy. Astron. Soc. \textbf{469}, 3928 (2016).

\bibitem{Mahmoodifar2013}S. Mahmoodifar and T. Strohmayer, Astrophys. J. \textbf{773}, 10 (2013).

\bibitem{Abbott2017a} B. P. Abbott et al. (LIGO Scientific Collaboration and Virgo Collaboration), Phys. Rev. Lett. \textbf{119}, 161101 (2017).

\bibitem{Abbott2017b} B. P. Abbott et al. (LIGO Scientific Collaboration and Virgo Collaboration), Astrophys. J. \textbf{848}, L12 (2017).

\bibitem{Zhou2018} E. P. Zhou, X. Zhou and A. Li, Phys. Rev. D \textbf{97}, 083015 (2018). 

\bibitem{Alford2012} M. G. Alford, S. Mahmoodifar and K. Schwenzer, AIP Conference Series \textbf{1492}, 257 (2012)

\bibitem{Bondarescu2009} R. Bondarescu, T. S. A., and I. Wasserman, Phys. Rev. D \textbf{79}, 104003 (2009).

\bibitem{Rezzolla2000} L. Rezzolla, F. K. Lamb, and S. L. Shapiro, Astrophys. J. \textbf{531}, L139 (2000).

\bibitem{cao2015} G. J. Cao, X. Zhou and N. Wang, Science China Physics, Mechanics \& Astronomy \textbf{58}, 5573 (2015).

\bibitem{Owen2010} B. J. Owen, Phys. Rev. D \textbf{82}, 104002 (2010).

\bibitem{Wette2008} K. Wette et al., Classical Quantum Gravity \textbf{25}, 235011 (2008).

\bibitem{Abadie2010} J. Abadie, et al., Astrophys. J. \textbf{722}, 1504 (2010).

\bibitem{ligo2019} B. P. Abbott, et al., arXiv: 1903.01901 (2019).

\bibitem{Punturo2010} M. Punturo, et al., Classical Quantum Gravity \textbf{27}, 194002 (2010). 

\bibitem{Li2018a} D. Li, et al., IEEE Microw. Mag. \textbf{19}, 112 (2018).

\bibitem{Wang2017} N. Wang, Sci. Sin-Phys .Mech. Astron. \textbf{47}, 7 (2017).

\bibitem{Watts2015} A. L. Watts, et al., Advancing Astrophysics with the Square Kilometre Array (AASKA14), 43 (2015).

\bibitem{Li2018} T. P. Li, et al., Sci. China Phys. Mech. Astron. \textbf{61}, 31011 (2018).

\bibitem{Keek2018} L. Keek, et al., Astrophys. J. \textbf{855}, L4 (2018).

\bibitem{Zhang2019} S. N. Zhang, et al., Sci. China Phys. Mech. Astron. \textbf{62}, 29502 (2019).
\end{thebibliography}

\end{document}